\documentclass[runningheads]{llncs}
\usepackage[T1]{fontenc}
\usepackage{array}
\usepackage{amssymb}
\usepackage{amsmath}
\usepackage{marvosym}
\usepackage{hyperref}

\usepackage{multirow}
\usepackage{graphicx}
\begin{document}

\title{Ariadne's Thread\protect\footnote{Ariadne's thread, the name comes from ancient Greek myth, tells of Theseus walking out of the labyrinth with the help of Ariadne's golden thread.} 
: Using Text Prompts to Improve Segmentation of Infected Areas from Chest X-ray images}

\titlerunning{Using Text Prompts to Improve Segmentation}
% If the paper title is too long for the running head, you can set
% an abbreviated paper title here
%

\author{Yi Zhong\and
Mengqiu Xu \and
Kongming Liang \and
Kaixin Chen \and
Ming Wu\textsuperscript{\Letter}
}

\authorrunning{Y. Zhong et al.}
% First names are abbreviated in the running head.
% If there are more than two authors, 'et al.' is used.
%
\institute{Beijing University of Posts and Telecommunications, China\\
\email{\{xiliang2017, xumengqiu, liangkongming, chenkaixin, wuming\}@bupt.edu.cn}
}
\maketitle              % typeset the header of the contribution

\begin{abstract}
Segmentation of the infected areas of the lung is essential for quantifying the severity of lung disease like pulmonary infections. Existing medical image segmentation methods are almost uni-modal methods based on image. 
%Due to the sensitivity and scarcity of healthcare data, 
However, these image-only methods tend to produce inaccurate results unless trained with large amounts of annotated data. To overcome this challenge, we propose a language-driven segmentation method that uses text prompt to improve to the segmentation result. Experiments on the QaTa-COV19 dataset indicate that our method improves the Dice score by 6.09\% at least compared to the uni-modal methods. Besides, our extended study reveals the flexibility of multi-modal methods in terms of the information granularity of text and demonstrates that multi-modal methods have a significant advantage over image-only methods in terms of the size of training data required.

\keywords{Multi-modal Learning \and Medical Image Segmentation }
\end{abstract}

\section{Introduction}
% %强调肺部感染区域的重要性，定性评估的依据，对于临床诊断和评估疾病有重要意义，在呼吸道大流行病（例如新冠）期间利用X光进行快速诊断和分级非常有意义。

% % 把related work中的部分提上来（对比学习需要大量配对数据）
% % LViT没有对文本特征进行深度编码，模型不够灵活
% % related work里直接讲医学中有local研究 然后讲LViT 再讲我们的方法

Radiology plays an important role in the diagnosis of some pulmonary infectious diseases, such as the COVID-19 pneumonia outbreak in late 2019\cite{yin2022sd}. With the development of deep learning, deep neural networks are more and more used to process radiological images for assisted diagnosis, such as disease classification, lesion detection and segmentation, etc. 
With the fast processing of radiological images by deep neural networks, some diagnoses can be obtained immediately, such as the classification of bacterial or viral pneumonia and the segmentation mask for pulmonary infections, which is important for quantifying the severity of the disease as well as its progression\cite{oulefki2021automatic}. Besides, these diagnoses given by the AI allow doctors to predict risks and prognostics in a "patient-specific" way\cite{mu2021progressive}. Radiologists usually take more time to complete lesion annotation than AI, and annotation results can be influenced by individual bias and clinical experience\cite{9098956}. Therefore, it is of importance to design automatic medical image segmentation algorithms to assist clinicians in developing accurate and fast treatment plans.\par

Most of the biomedical segmentation methods\cite{cao2023swin,hatamizadeh2022swin,zhou2019unet++,DBLP:journals/corr/abs-1804-03999,hatamizadeh2022unetr} are improved based on U-Net\cite{ronneberger2015u}. However, the performance of these image-only methods is constrained by the training data, which is also a dilemma in the medical image field. Radford et al. proposed CLIP\cite{radford2021learning} in 2021, where they used 4M image-text pairs for contrastive learning. With the rise of multi-modal learning in the recent years, there are also methods\cite{wang2022cris,rao2022denseclip,bhalodia2021improving,muller2022radiological,li2022lvit} that focus on vision-language pretraining/processing and applying them on local tasks. Li et al. proposed a language-driven medical image segmentation method LViT\cite{li2022lvit}, using a hybrid CNN-Transformer structure to fuse text and image features. However, LViT uses an early fusion approach and the information containd in the text is not well represented. In this paper, we propose a multi-modal segmentation method that using independent text encoder and image encoder, and design a GuideDecoder to fuse the features of both modalities at decoding stage. Our main contributions are summarized as follow:
\begin{itemize}
\item We propose a language-driven segmentation method for segmenting infected areas from lung x-ray images. Source code of our method see: \par
\href{https://github.com/Junelin2333/LanGuideMedSeg-MICCAI2023}{https://github.com/Junelin2333/LanGuideMedSeg-MICCAI2023}
\item The designed GuideDecoder in our method can adaptively propagate sufficient semantic information of the text prompts into pixel-level visual features, promoting consistency between two modalities.
\item We have cleaned the errors contained in the text annotations of QaTa-COV19\cite{degerli2022osegnet} and contacted the authors of LViT to release a new version.
\item Our extended study reveals the impact of information granularity in text prompts on the segmentation performance of our method, and demonstrates the significant advantage of multi-modal method over image-only methods in terms of the size of training data required.
\end{itemize}

\section{Method}

% %先写一个总括  概括方法的主要内容 method里的两个子标题  要讨论

% \subsection{Using Text Prompts to Improve Segmentation Performance}

\begin{figure}
    \centering
    \includegraphics[width=\textwidth]{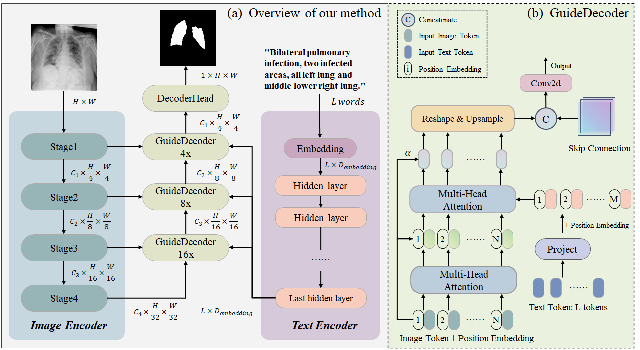}
    \caption{\textbf{The overview of the our proposed method (a) and the detail of the GuideDecoder in out method. }Our proposed approach uses a modular design where the model consists mainly of an image encoder, a text encoder and several GuideDecoders. The GuideDecoder is used to adaptively propagate semantic information from textual features to visual features and output decoded visual features.} 
    \label{overview}
\end{figure}

The overview of our proposed method is shown in Fig.~\ref{overview}(a). The model consists of three main components: Image Encoder, Text Encoder and GuideDecoder that enables multi-modal information fusion. As you can see, our proposed method uses a modular design. Compared to early stage fusion in LViT, our proposed method in modular design is more flexible. For example, when our method is used for brain MRI images, thanks to the modular design, we could first load pre-trained weights trained on the corresponding data to separate visual and text encoders, and then only need to train GuideDecoders. 
% The modular design significantly reduces training costs, which is difficult to achieve with LViT.

\subsubsection{Visual Encoder \& Text Encoder} The Visual Encoder used in the model is ConvNeXt-Tiny\cite{liu2022convnet}. For an input image $I\in \mathbb{R}^{H\times W\times1}$, we extract multiple visual features from the four stages of ConvNeXt-Tiny, which are defined as $f_4\in\mathbb{R}^{\frac{H}{4}\times\frac{W}{4}\times C_{1}}$, $f_8\in\mathbb{R}^{\frac{H}{8}\times\frac{W}{8}\times C_{2}}$,
$f_{16}\in\mathbb{R}^{\frac{H}{16}\times\frac{W}{16}\times C_{3}}$ and 
$f_{32}\in\mathbb{R}^{\frac{H}{32}\times\frac{W}{32}\times C_{4}}$,
Note that $C$ is the feature dimension, $H$ and $W$ are the height and width of the original image.
For an input text prompt $T \in \mathbb{R}^L $, We adopt the CXR-BERT\cite{boecking2022making} to extract text features $g_t \in \mathbb{R}^{L\times C}$. Note that $C$ is the feature dimension, $L$ is the length of the text prompt.

\subsubsection{GuideDecoder}
Due to our modular design, visual features and textual features are encoded independently by different encoders. Therefore, the design of the decoder is particularly important, as we can only fuse multi-modal features from different encoders in post stage. The structure of GuideDecoder is shown in Fig.~\ref{overview}(b). The GuideDecoder first processes the input textual features and visual features before performing multi-modal interaction.  

The input textual features first go through a projection module (i.e. Project in the figure) that aligns the dimensionality of the text token with that of the image token and reduces the number of text tokens. The projection process is shown in Equation 1. 
\begin{equation}
 f_t = \sigma(Conv(T W_T))
\end{equation}
where $W_T$ is a learnable matrix, $Conv(\cdot)$ denotes a $1\times1$ convolution layer, and $\sigma(\cdot)$ denotes the ReLU activation function. Given an input feature $T \in \mathbb{R}^{L\times D} $, the output projected features is $f_t \in \mathbb{R}^{M \times C_1}$, where $M$ is the number of tokens after projection and $C_1$ is the dimension of the projected features, consistent with the dimension of the image token. \par
For the input visual features $I\in \mathbb{R}^{H\times W\times C_1}$, after adding the position encoding we use self-attention to enhance the visual information in them to obtain the evolved visual features. The process is shown in Equation 2. 
\begin{equation}
    f_i = I + LN(MHSA(I))
\end{equation}
where $MHSA(\cdot)$ denotes Multi-Head Self-Attention layer, $LN(\cdot)$ denotes Layer Normalization, and finally the evolved visual features $f_i \in \mathbb{R}^{H\times W\times C_1}$ with residuals could be obtained. 

After those, the multi-head cross-attention layer is adopted to propagate fine-grained semantic information into the evolved image features. To obtain the multi-modal feature $f_c \in \mathbb{R}^{H\times W\times C_1}$, the output further computed by layer normalization and residual connection:
\begin{equation}
    f_c = f_i + \alpha (LN(MHCA(f_i,f_t)))
\end{equation}
where $MHCA(\cdot)$ denotes multi-head cross-attention and $\alpha$ is a learnable parameter to control the weight of the residual connection.

Then, the multi-modal feature $f_c \in \mathbb{R}^{(H\times W)\times C_1} $ would be reshaped and upsampling to obtain $f'_c \in \mathbb{R}^{H'\times W'\times C_1}$. Finally the $f'_c$ is concatenated with $f_s\in \mathbb{R}^{H'\times W'\times C_2}$ on the channel dimension, where $f_s$ is the low-level visual feature obtained from visual encoder via skip connection. The concatenated features are processed through a convolution layer and a ReLU activation function to obtain the final decoded output $f_o \in \mathbb{R}^{H'\times W'\times C_2}$
\begin{gather}
    f'_c = Upsample(Reshape(f_c)) \notag \\
    f_o = \sigma(Conv([f'_c, f'_s]))
\end{gather}
where 
% $\sigma(\cdot)$ represents the ReLU activation function and 
$[\cdot,\cdot]$ represents the concatenate operation on the channel dimension.

% % 写不下了
% %\subsection{Generating text prompts from images using supervised learning and categorical coding prediction.}

\section{Experiments}

\subsection{Dataset}

The dataset used to evaluate our method performance is the QaTa-COV19 dataset\cite{degerli2022osegnet}, which is compiled by researchers from Qatar University and Tampere University. It consists of 9258 COVID-19 chest radiographs with pixel-level manual annotations of infected lung areas, of which 7145 are in the training set and 2113 in the test set. However, the original QaTa-COV19 dataset does not contain any matched text annotations. 

Li et al. \cite{li2022lvit}have made significant contributions by extending the text annotations of the dataset, their endeavors are worthy of commendation. We conducted a revisitation of the text annotations and found several notable features. Each sentence consists of three parts, containing position information at different granularity. However, these sentences cannot be considered as medical reports for lacking descriptions of the disease, we consider them as a kind of "text prompt" just as the title of the paper states.
% For example, the text annotation of id 'covid\_1000.png' image is \emph{"Bilateral pulmonary infection, two infected areas, lower left lung and upper middle lower right lung."}, the first part refers to unilateral or bilateral lung infection, the second part refers to the number of infected areas and the third part refers to the location of the infected areas (upper, middle, lower in the left lung and right lung, six areas in total).
% We will further discuss the impact of introducing different fine-grained information on segmentation performance in our extended studies. \par

% % 这里补一个图 image mask text（框出来指代什么粒度）
% % 再补一个表，列举数据集中出现的错误种类和错误代表
Besides, we found some obvious errors (e.g. misspelled words, grammatical errors and unclear referents) in the extended text annotations. We have fixed these identified errors and contacted the authors of LViT to release a new version of the dataset. Dataset see Github link: \href{https://github.com/HUANGLIZI/LViT}{https://github.com/HUANGLIZI/LViT}

\subsection{Experiment Settings}
% 包括数据集的划分、超参数设置、评估指标、实验环境
Following the file name of the subjects in the original train set, we split the training set and the validation set uniformly in the ratio of 80\% and 20\%. Therefore, the training set has a total of 5716 samples, the validation set has 1429 samples and the test set has 2113 samples. 
% We use the training set to train network parameters, use the validation set to select the best model, and use the test set for final evaluation.
All images are cropped to $224\times224$ and the data is augmented using a random zoom with 10\% probability.

We used a number of open source libraries including but not limited to PyTorch, MONAI\cite{cardoso2022monai} and Transformers\cite{wolf-etal-2020-transformers} to implement our method and baseline approach. We use PyTorch Lightning for the final training and inference wrapper. All the methods are training on one NVIDIA Tesla V100 SXM3 32GB VRAM GPU. We use the Dice loss plus Cross-entropy loss as the loss function, and train the network using AdamW optimization with a batch size of 32. We utilize the cosine annealing learning rate policy, the initial learning rate is set to 3e-4 and the minimal learning rate is set to 1e-6. 
%The maximum number of training epochs is set to 200, and we also set an early stop policy, which automatically stops the training when the performance of the model does not improve after 20 epochs. 
%As the library 'bert\_embedding' used in the LViT source code is obsolete and no longer maintained, we have changed the LViT related code to use the transformers library and the CXR-BERT from the Hugging Face community as the word embedding in LViT, in line with our proposed method.
% 缺一个公式 dice loss + cross entropy loss

We used three metrics to evaluate the segmentation results objectively: Accuracy, Dice coefficient and Jaccard coefficient. %Dice coefficient and Jaccard coefficient are the two most representative commonly used evaluation metrics for medical image segmentation.
Both Dice and Jaccard coefficient calculate the intersection regions over the union regions of the given predicted mask and  ground truth, where the Dice coefficient is more indicative of the segmentation performance of small targets.

\subsection{Comparison Experiments}

We compared our method with common mono-modal medical image segmentation methods and with the LViT previously proposed by Li et al. The quantitative results of the experiment are shown in Table~\ref{comparison}. UNet$++$ achieves the best performance of the mono-modal approach. Comparing to UNet$++$, our method improves accuracy by 1.44\%, Dice score by 6.09\% and Jaccard score by 9.49\%. Our method improves accuracy by 1.28\%, Dice score by 4.86\% and Jaccard coefficient by 7.66\% compared to the previous multi-modal method LViT. In general, using text prompts could significantly improve segmentation performance.\par

\begin{table}[]
\caption{\textbf{Comparisons with some mono-modal methods and previous multi-modal method on QaTa-COV19 test set. }'$*$' denotes these methods use text prompt and CXR-BERT as the text embedding encoder.} %In the table header, \textbf{Acc} indicates accuracy, \textbf{Dice} indicates Dice Score and \textbf{Jaccard} indicates Jaccard Coefficient.}
\renewcommand\arraystretch{1.05}
\centering
\begin{tabular}{m{4cm}<{\centering}|m{3cm}<{\centering}|m{1.5cm}<{\centering}|m{1.5cm}<{\centering}|m{1.5cm}<{\centering}}
\hline
\multicolumn{2}{c|}{\textbf{Method}}   & \textbf{Acc}   & \textbf{Dice}  & \textbf{Jaccard}   \\ \hline
\multicolumn{1}{c|}{\multirow{4}{*}{  Mono-Modal  }}  & Unet           & 0.9584          & 0.8299          & 0.7092          \\
\multicolumn{1}{c|}{}                             & Unet++         & \textbf{0.9608} & \textbf{0.8369} & \textbf{0.7196} \\
\multicolumn{1}{c|}{}                             & Attention Unet & 0.9567          & 0.8240          & 0.7006          \\
\multicolumn{1}{c|}{}                             & Swin UNETR     & 0.9511          & 0.8002          & 0.6669          \\ \hline
\multicolumn{1}{c|}{\multirow{2}{*}{  Multi-Modal*  }} & LViT           & 0.9624          & 0.8492          & 0.7379          \\
\multicolumn{1}{c|}{}                             & Our Method     & \textbf{0.9752} & \textbf{0.8978} & \textbf{0.8145} \\ \hline
\end{tabular}
\label{comparison}
\end{table}

\begin{figure}
\centering
\includegraphics[width=\textwidth]{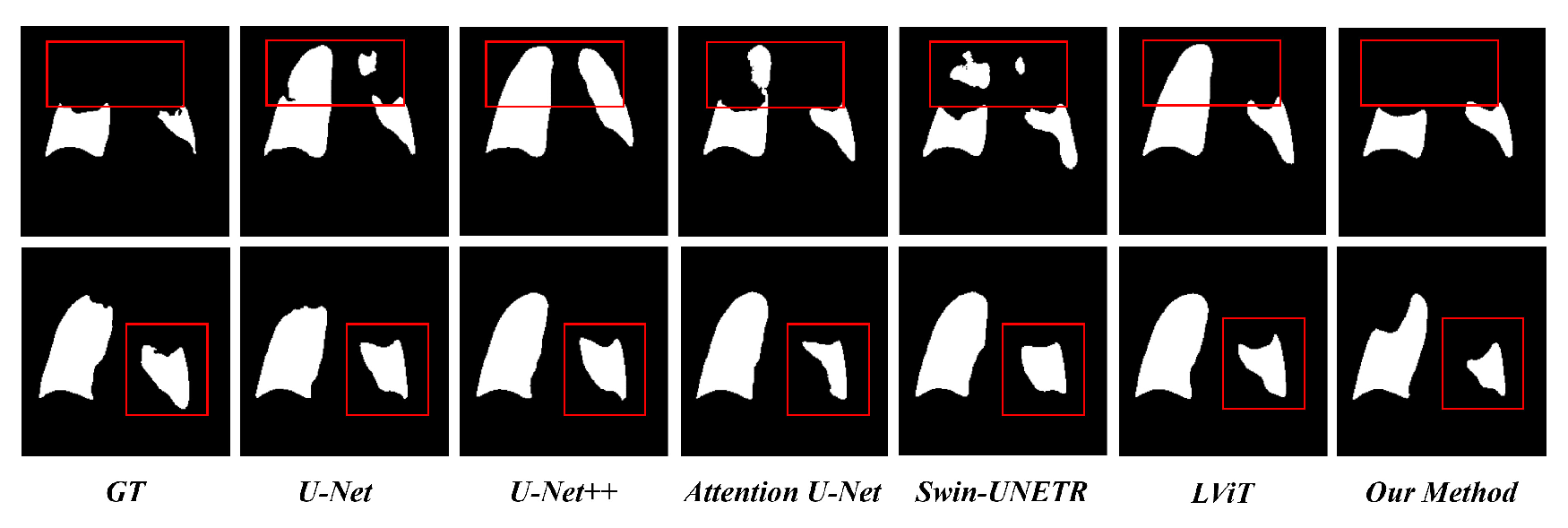}
    \caption{\textbf{Qualitative results on QaTa-COV19.}The sample image in the first row is from 'sub-S09345\_ses-E16849\_run-1\_bp-chest\_vp-ap\_dx.png' and The sample image in the second row is from 'sub-S09340\_ses-E17282\_run-1\_bp-chest\_vp-ap\_cr.png'.}
    \label{qualitative}
\end{figure}

The results of the qualitative experiment are shown in Fig.~\ref{qualitative}. The image-only mono-modal methods tend to generate some over-segmentation, while the multi-modal approach refers to the specific location of the infected region through text prompts to make the segmentation results more accurate. 
% The results of the qualitative analysis also show that the introduction of extra text information could guide the model to generate more accurate segmentation maps.

\subsection{Ablation Study}
Our proposed method introduces semantic information of text in the decoding process of image features and designs the GuideDecoder to let the semantic information in the text guide the generation of the final segmentation mask. We performed an ablation study on the number of GuideDecoder used in the model and the results are shown in the Table~\ref{ablation}. 

\begin{table}[]
\caption{\textbf{Ablation studies on QaTa-COV19 test set. }We used different numbers($0\sim3$) of GuideDecoders in the model to verify the effectiveness of the GuideDecoder. Note that the GuideDecoder in the model is replaced in turn by the Decoder in the UNet, 'w/o text' means without text and the model use UNet Decoders only.}
\begin{center}
\renewcommand\arraystretch{1.05}
\begin{tabular}{m{2.5cm}<{\centering}|m{1.5cm}<{\centering}|m{1.5cm}<{\centering}|m{1.5cm}<{\centering}}
\hline
\textbf{Method} & \textbf{Acc} & \textbf{Dice} & \textbf{Jaccard} \\ \hline
w/o text        & 0.9610       & 0.8414        & 0.7262                 \\
1 layer         & 0.9735       & 0.8920        & 0.8050                 \\
2 layers        & 0.9748       & 0.8963        & 0.8132                 \\
3 layers        & \textbf{0.9752}       & \textbf{0.8978}        & \textbf{0.8144}   \\ \hline
\end{tabular}
\end{center}
\label{ablation}
\end{table}
As can be seen from the Table~\ref{ablation}, the segmentation performance of the model improves as the number of GuideDecoders used in the model increases. The effectiveness of GuideDecoder could be proved by these results.

\subsection{Extended Study}

Considering the application of the algorithm in clinical scenarios, we conducted several interesting extension studies based on the QaTa-COV19 dataset with the text annotations. It is worth mentioning that the following extended studies were carried out on our proposed method.

\subsubsection{Impact of text prompts at different granularity on segmentation performance.}
\begin{figure}[]
\centering
\includegraphics[width=\textwidth]{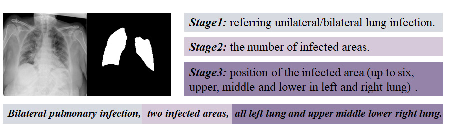}
\caption{\textbf{The Split Example of Different Stages in Text Annotation. }The chest x-ray, mask and text annotation correspond to 'coivd\_1.png'. We divide the sentence into three stages and distinguished them with a background colour, the darker the background colour the more detailed the stage has position information.} 
\label{stage example}
\end{figure}

In section 3.1 we mention that each sample is extended to a text annotation with three parts containing positional information at different granularity, as shown in the Fig.~\ref{stage example}. Therefore we further explored the impact of text prompts at different granularity on segmentation performance of our method and the results are shown in Table~\ref{prompt stage}. 

% 原先准备的图放到这里来，4.1就不讲了

% [xmq]
% \begin{table}[]
% \begin{center}
% \renewcommand\arraystretch{1.25}
% \resizebox{\linewidth}{!}{
% \begin{tabular}{m{2.25cm}<{\centering}|m{6.5cm}<{\centering}|m{1.25cm}<{\centering}|m{1.25cm}<{\centering}|m{1.25cm}<{\centering}}
% \hline
% \textbf{Stage of Text Prompt} & \textbf{Example of Text} & \textbf{Acc} & \textbf{Dice} & \textbf{Jaccard} \\ \hline
% w/o text        & None                                               & 0.9610 & 0.8414 & 0.7262 \\ \hline
% stage1 + stage2   & Bilateral pulmonary infection, two infected areas. & 0.9648 & 0.8557 & 0.7479 \\ \hline
% stage3          & Lower left lung and lower right lung.  & \textbf{0.9753} & \textbf{0.8981} & \textbf{0.8151} \\ \hline
% stage1 + stage2 + stage3 & Bilateral pulmonary infection, two infected areas,  lower left lung and lower right lung. &
%   0.9752 &
%   0.8978 &
%   0.8145 \\ \hline
% \end{tabular}}
% \end{center}
% \end{table}

\begin{table}[]
\caption{\textbf{Study of text prompts at different granularity and segmentation performance. }The term \textit{w/o text} in the table means \textit{without text}, while \textit{Stage1}, \textit{Stage2}, and \textit{Stage3} represent the three parts of each text annotation. See Fig.~\ref{stage example} for examples of the different Stages.}
\centering
\begin{tabular}{m{4.5cm}<{\centering}|m{1.25cm}<{\centering}|m{1.25cm}<{\centering}|m{1.25cm}<{\centering}}
\hline
\textbf{Stage of Text Prompt} & \textbf{Acc} & \textbf{Dice} & \textbf{Jaccard} \\ \hline
w/o text                      & 0.9610       & 0.8414        & 0.7262           \\
stage1 + stage2               & 0.9648       & 0.8557        & 0.7479           \\
stage3                        & \textbf{0.9753} & \textbf{0.8981}  & \textbf{0.8151}   \\
stage1 + stage2 + stage3      & 0.9752       & 0.8978        & 0.8145           \\ \hline
\end{tabular}
\label{prompt stage}
\end{table}

%\textit{Stage1} corresponds to the first part of the sentence, which indicates the coarsest-grained position information, i.e., unilateral/bilateral lung infection. \textit{Stage2} corresponds to the second part of the sentence, which describes the number of infected areas. \textit{Stage3} corresponds to the third part of the sentence, which refers to the most detailed position information, i.e., the upper, middle, and lower parts of the left and right lungs. The results in the table show that the segmentation performance of our proposed method is driven by the fine-grained position information contained in the text prompt. 

The results in the table show that the segmentation performance of our proposed method is driven by the granularity of the position information contained in the text prompt. Our proposed method achieved better segmentation performance when given a text prompt with more detailed position information. 
Meanwhile, we observed that the performance of our method is almost identical when using two types of text prompts, i.e. \textit{Stage3} alone and \textit{Stage1 + Stage2 + Stage3}. It means the most detailed position information in the text prompt plays the most significant role in improving segmentation performance. But this does not mean that other granularity of position information in the text prompt does not contribute to the improvement in segmentation performance. Even when the input text prompts contain only the coarsest location information (\textit{Stage1 + Stage2} items in the Table~\ref{prompt stage}), our proposed method yielded a 1.43\% higher Dice score than the method without text prompt. %Therefore, our method allows for some flexibility in the granularity of the information in the text prompt.

%下面这个子标题里面是用量（amount）还是数量（number）比较好？ 多了一个选择 quantity
\subsubsection{Impact of the size of training data on segmentation performance.}
As shown in Table~\ref{size}, our proposed method demonstrates highly competitive performance even with a reduced amount of training data. With only a quarter of the training data, our proposed method achieves a 2.69\% higher Dice score than UNet++, which is the best performing mono-modal model trained on the full dataset. This provides sufficient evidence for the superiority of multi-modal approaches and the the fact that suitable text prompts could significantly help improve the segmentation performance.

\begin{table}[]
\centering
\caption{\textbf{Study of the size of training data and segmentation of performance.} Note that the performance of UNet++ is used as a comparative reference. }
\begin{tabular}{m{5.25cm}<{\centering}|m{1.25cm}<{\centering}|m{1.25cm}<{\centering}|m{1.25cm}<{\centering}}
\hline
\textbf{Method}                        & \textbf{Acc} & \textbf{Dice} & \textbf{Jaccard} \\ \hline
Unet++ (using 10\% training data)  & 0.9425            & 0.7706        & 0.6268               \\
Ours (using 10\% training data)  & 0.9574            & 0.8312        & 0.7111                 \\
Unet++ (using 100\% training data)     & 0.9608            & 0.8369        & 0.7196     \\
Ours (using 15\% training data)  & 0.9636            & 0.8503        & 0.7395                 \\
Ours (using 25\% training data)  & 0.9673            & 0.8638        & 0.7602                 \\
Ours (using 50\% training data)  & 0.9719            & 0.8821        & 0.7891                 \\
Ours (using 100\% training data) & \textbf{0.9752}  & \textbf{0.8978}  & \textbf{0.8145}  \\ \hline
\end{tabular}
\label{size}
\end{table}

We observed that when the training data was reduced to 10\%, our method only began to exhibit inferior performance compared to UNet++, which was trained with all available data. Similar experiments could be found in the LViT paper. Therefore, it can be argued that multi-modal approaches require only a small amount of data (less than 15\% in the case of our method) to achieve performance equivalent to that of mono-modal methods.

\section{Conclusion}

In this paper, we propose a language-driven method for segmenting infected areas from lung x-ray images. The designed GuideDecoder in our method can adaptively propagate sufficient semantic information of the text prompts into pixel-level visual features, promoting consistency between two modalities. The experimental results on the QaTa-COV19 dataset indicate that the multi-modal segmentation method based on text-image could achieve better performance compared to the image-only segmentation methods. Besides, we have conducted several extended studies on the information granularity of the text prompts and the size of the training data, which reveals the flexibility of multi-modal methods in terms of the information granularity of text and demonstrates that multi-modal methods have a significant advantage over image-only methods in terms of the size of training data required.

\subsubsection{Acknowledgements} This work was supported by NSFC under Grant 62076093 and MoE-CMCC "Artifical Intelligence" Project No. MCM20190701.

\newpage

% For citations of references, we prefer the use of square brackets
% and consecutive numbers. Citations using labels or the author/year
% convention are also acceptable. The following bibliography provides
% a sample reference list with entries for journal
% articles~\cite{ref_article1}, an LNCS chapter~\cite{ref_lncs1}, a
% book~\cite{ref_book1}, proceedings without editors~\cite{ref_proc1},
% and a homepage~\cite{ref_url1}. Multiple citations are grouped
% \cite{ref_article1,ref_lncs1,ref_book1},
% \cite{ref_article1,ref_book1,ref_proc1,ref_url1}.

%
% ---- Bibliography ----
%
% BibTeX users should specify bibliography style 'splncs04'.
% References will then be sorted and formatted in the correct style.
%
\bibliographystyle{splncs04}
\bibliography{reference}

\end{document}